# Using Cardinality Matching to Design Balanced and Representative Samples for Observational Studies

Bijan A. Niknam, BS[1,2]
Jose R. Zubizarreta, PhD[1,2,3,4]

[1] Department of Health Care Policy, Harvard Medical School
[2] CAUSALab, Harvard TH Chan School of Public Health
[3] Department of Statistics, Harvard University
[4] Department of Biostatistics, Harvard TH Chan School of Public Health

**Address Correspondence to:** Jose R. Zubizarreta, PhD, Associate Professor, Harvard University, 180A Longwood Ave Office 307-D, Boston, MA 02115. (T) (617) 432-1266; (F) (617) 432-0173; Email: zubizarreta@hcp.med.harvard.edu.

**Acknowledgements:** This work was supported through a Patient-Centered Outcomes Research Institute (PCORI) Project Program Award (ME-2019C1-16172). PCORI had no role in the design and conduct of the study; collection, management, analysis, and interpretation of the data; preparation, review, or approval of the manuscript; and decision to submit the manuscript for publication. For graphics, we thank Xavier Alemany, BS.

**If you wish to cite this work, please use the following citation:**
Niknam BA, Zubizarreta JR. Using Cardinality Matching to Design Balanced and Representative Samples for Observational Studies. *JAMA*. 2022;327(2):173–174. doi:10.1001/jama.2021.20555.

**Link to published version:** https://jamanetwork.com/journals/jama/fullarticle/2787922

Cardinality matching is a computational method for finding the largest possible number of matched pairs of exposed and unexposed individuals from an observational dataset, with specified patterns of baseline characteristics that represent a target population for analysis. In a report in *JAMA Network Open*, Benjet et al. used cardinality matching to examine the association of neighborhood exposure to violence during an armed conflict in Nepal with the incidence of major depressive disorder in younger children and older individuals.(1) The authors found that children under age 11 at the start of the conflict who were exposed to violence during the conflict were significantly more likely to develop major depressive disorder than matched children residing in unexposed neighborhoods. In contrast, there was no association between exposure to violence and development of depression among individuals aged 11 or older.

**Use of the Method**

*Why is Cardinality Matching Used?*

In an observational study, treatment assignment is not controlled by the investigator, often resulting in exposed and unexposed groups that are not comparable. This makes evaluating the possible effects related to the exposure challenging because differences in outcomes could be due to differences in characteristics other than the exposure. Randomized studies overcome this problem by controlling treatment assignment and producing treated and control groups that are similar on average. In many contexts, such as studying the effects of violence, randomization is infeasible. Thus, an alternative is to analyze observational data in such a way as to attempt to approximate a randomized experiment as closely as possible.(2)

An important component of designing observational studies is to take steps so that exposed and unexposed groups being compared are as similar as possible with respect to other characteristics (i.e., confounders) that might be related to the exposure and that also might influence outcomes. Matching is a method commonly used to construct comparable or "balanced" exposed and unexposed groups from observational data. Often, this is done by first estimating each participant's propensity score, which is their probability of having the exposure under study based on their characteristics.(3,4) Then, exposed participants are matched to unexposed participants with similar propensity scores,(5,6) often using a nearest-neighbor algorithm that finds the closest matches first, proceeding until the differences in propensity scores become too large to be considered matched.

After matching, the balance of measured characteristics can be checked transparently through simple figures and tables, and if the characteristics are not balanced, the matches can be modified until they are. However, this method often excludes some exposed participants because they may be too different from any of the unexposed participants to be matched. Once exposed participants are excluded, the study sample no longer captures the population of all exposed participants. Hence, it is desirable to perform matching in a way that maximizes the cardinality, or sample size, of the matched study and that results in study groups that are balanced and representative of the entire target population. With cardinality matching, investigators can include as many exposed participants as possible. In the study by Benjet et al., the matches included every child and older individual in the dataset who lived in neighborhoods exposed to violence.(1)

*Description of Cardinality Matching*

Cardinality matching simultaneously addresses the concerns of balance, sample size, and representativeness of the matched sample. The method uses optimization techniques to find the largest possible matched sample that is balanced relative to a description (or profile) of a target population specified by the investigator before matching.(7) This helps to ensure that the matched sample has a profile of characteristics similar to the target population. Cardinality matching also balances potentially confounding characteristics directly and can practically do so in large datasets comprising hundreds of thousands of observations.(8)

The **Figure** illustrates the process of cardinality matching. At left is the overall target population, followed by the exposed and unexposed samples observed by the investigator. Because treatment assignment was not random, these samples are dissimilar with respect to the observed characteristics of age, gender, and residence in a neighborhood with both a school and health clinic nearby (denoted by red outlines). If the investigator wanted to assess the association between a treatment and an outcome in the overall population, the first step involves specifying a profile that describes the characteristics of that population. Then, the algorithm finds the largest possible matched sample of exposed and unexposed participants that is representative of the target population. After the balanced sample is found, the selected participants are matched to form pairs that are homogeneous on their characteristics.

The target profile is flexible. For example, if the study was intended to understand how the findings reported by Benjet et al. would generalize to children in another region of Nepal, the investigator could set a profile corresponding to the other region's children and their neighborhoods, and then find the largest sample of paired exposed and unexposed children who match the characteristics of the target population. Thus, the investigator could gain insight into the likely association between exposure to violence and mental health outcomes among children in the other region (e.g., due to differences in age distributions). Moreover, direct access to individual-level data from the other region would not be necessary to set the target profile; if aggregate information was publicly available, the investigator could use that instead.

**Limitations of the Method**

In observational studies, methods such as cardinality matching are helpful to adjust for differences in characteristics that are measured, but not characteristics that remain unmeasured or are unknown. Thus, it remains possible that unobserved confounders are responsible for observed associations rather than exposure to the treatment. Therefore, investigators must be careful to include as many relevant observable characteristics as possible.

To be valid, extrapolation of associations with exposures from observed data onto other populations relies on the assumption that the relationships between baseline characteristics, exposure, and outcomes are the same for both the observed data and the target populations. In practice, this assumption is often difficult to test or verify.

Although cardinality matching constructs the largest possible matched sample, it may be less statistically efficient or reduce the power of a study relative to alternative approaches such as weighting.(9) With weighting, each exposed and unexposed participant is given more or less emphasis, such that observed characteristics are balanced between exposed and unexposed groups. While weighting may prioritize statistical efficiency, matching generally prioritizes study interpretability. Specifically, even though weighting may include more participant information in the study by allowing flexible weights, in cardinality matching the unit of analysis (e.g., the participant) is unchanged, making outcome analysis and reporting of results easier. Another option is regression modeling; however, this approach is sensitive to additional statistical assumptions.

**How was the Method used in this study?**

The study by Benjet et al. (1) examined a dataset comprised of children and older individuals who were living in 151 neighborhoods in the Chitwan Valley in Nepal during an armed conflict that spanned 2000 to 2006. Fifteen of these neighborhoods were exposed to violence, defined as the occurrence of at least two beatings within 1 km of the neighborhood. Collectively, 520 individuals (mean age at end of conflict, 23.6 years) resided in exposed neighborhoods, including 197 younger than age 11 at the start of the conflict (mean age at end of conflict, 11.4 years). The authors used cardinality matching to identify 520 similar individuals who were living in unexposed neighborhoods. To understand associations by age, the researchers matched individuals exactly by age category (younger than 11 vs. 11 or older) and, because the conflict was along ethnic lines, they also matched exactly by ethnicity. Additionally, the authors balanced the comparison groups with respect to the means of individuals' gender, education level, and characteristics of their neighborhoods, to try to ensure that their environments were also comparable.

After matching, the authors examined the outcome of major depressive disorder, and found that 25 (12.69%) of 197 children younger than age 11 who were exposed to violence developed major depressive disorder later in life (ages 15 to 29 at the time of outcome assessment), compared to 5.08% of matched unexposed children, a statistically significant difference ($P = 0.008$). In contrast, the match of exposed and unexposed individuals aged 11 or older showed no significant difference in subsequent development of major depressive disorder (7.74% vs. 6.81%, $P = 0.65$).

**How should the Method be interpreted?**

Benjet et al. used cardinality matching to examine the association between individuals' exposure to violence during an armed conflict in Nepal and their subsequent development of mental health disorders, and documented a significantly higher rate of major depressive disorder among exposed younger children. In observational studies, cardinality matching can be used to approximate a randomized experiment, enhancing internal validity by balancing observed characteristics transparently, while also strengthening external validity by balancing toward a target population. Like in any observational study, researchers using cardinality matching should carefully consider the assumptions and measurements necessary to find strong evidence of the effects of treatments, interventions, or exposures.

**Figure. Building Matched Samples that are Balanced and Representative by Design.** At left is the target population. Next are the observed samples of dissimilar exposed and unexposed study participants. After matching toward the target population, the matched exposed and unexposed participants are very similar on their characteristics, and also very similar to the target population. The balanced samples are matched to form pairs that are homogenous on their characteristics.

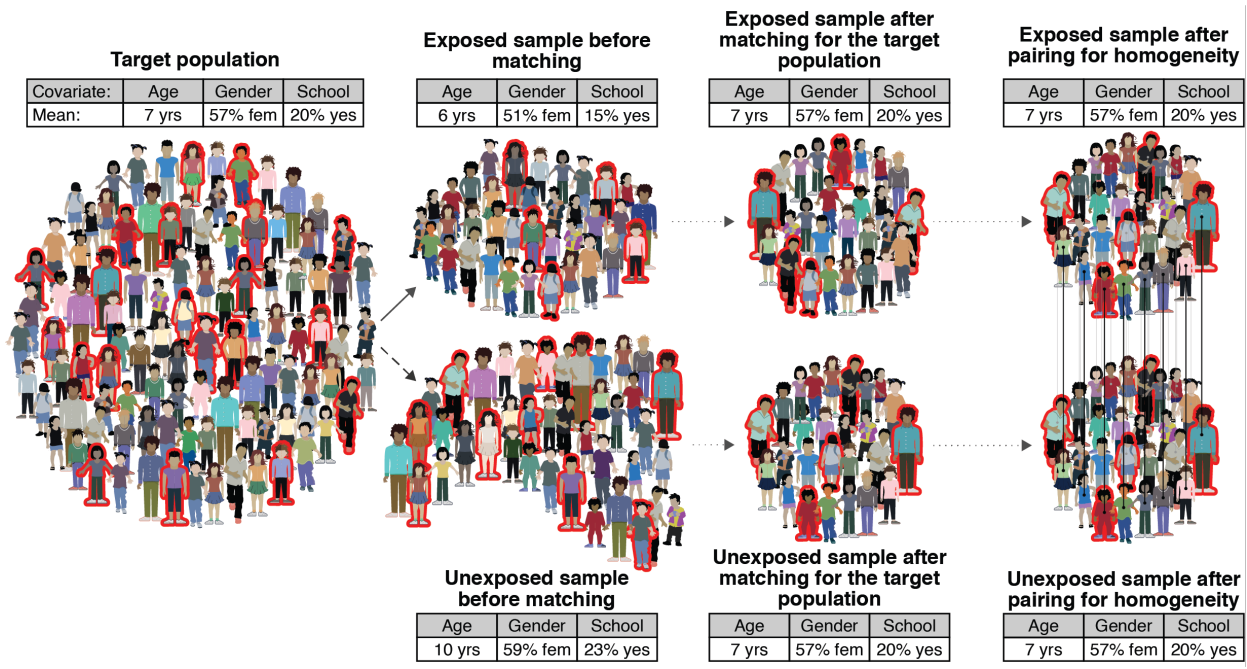